# Towards 250-m gigabits-per-second underwater wireless optical communication using a low-complexity ANN equalizer

Xiaohe Dong,[1,3] Kuokuo Zhang,[1,3] Caiming Sun,[1,*] Jun Zhang,[2] Aidong Zhang,[1] and Lijun Wang[2]

[1]*Shenzhen Institute of Artificial Intelligence and Robotics for Society (AIRS), School of Science and Engineering, The Chinese University of Hong Kong, Shenzhen 518172, Guangdong, China*
[2]*Key Laboratory of Luminescence Science and Technology, Chinese Academy of Sciences & State Key Laboratory of Luminescence and Applications, Changchun Institute of Optics, Fine Mechanics and Physics Chinese Academy of Sciences, Changchun 130033, China*
[3]*These authors contributed equally: Xiaohe Dong, Kuokuo Zhang*
*cmsun@cuhk.edu.cn

**Abstract:** The breakthroughs of communication distance and data rate have been eagerly anticipated by scientists in the area of underwater wireless optical communication (UWOC), which is seriously limited by the obvious aquatic attenuation in underwater channel. High-power laser source and ultra-sensitive photodetector are straightforward to extend the UWOC distance. However, nonlinear impairments caused by bandwidth-limited high-power transmitter and sensitive receiver severely degrade the data rate of long-distance UWOC. In this paper, we develop a UWOC system using a high-power transmitter by beam combining of 8-channel cascaded laser diodes (LD) and a sensitive receiver by a silicon photomultiplier (SiPM). The combined linear equalizer and low-complexity Artificial Neural Network (ANN) equalizer are used to achieve 1-Gbps data transmission over a 250-m UWOC system. To the best of our knowledge, this is the first Gbps-level UWOC experimental demonstration in >250-meter underwater transmission that has ever been reported. To lower the complexity of the ANN equalizer, a linear equalizer is applied first in order to prune the input size of the ANN equalizer. The optimal input size of the ANN equalizer is identified as 9. And the ANN architecture consists of two hidden layers, with 10 neurons in the first layer and a single neuron in the second layer. The performance of the proposed ANN-based system is compared with that of systems employing Volterra and linear equalizers. The bit error rate (BER) at data rate of 1 Gbps over a 250-m UWOC is reduced to $3.4\times10^{-3}$ with the combined linear and ANN equalizer, which is below the hard-decision forward error correction (HD-FEC) limit. In contrast, the linear and Volterra equalizer-based systems achieve data rates of 500 Mbps and 750 Mbps, respectively.

## 1. Introduction

The ongoing ocean exploration, such as oceangraphic research, subsea resource development, ocean environmental surveillance, underwater robot and so on, has accelerated the development of high-speed underwater wireless communication [1-6]. Underwater wireless optical communication (UWOC) has attractive merits of high speed and low latency due to the low attenuation "window" in ocean water, compared to underwater acoustic or radio frequency (RF) communication counterparts [4, 7]. Long-distance and high-speed UWOC has attracted plenty of research efforts among more and more scientists. However, to date, the communication distance and speed of UWOC demonstrated by experiments or commercial products still lag far behind the theoretical prediction by Monte Carlo numerical simulations [8], which forecasts that the communication distance can reach 500 m in ocean water. Table 1 summarizes the representative setup and performance of recent UWOC systems according to the transmission distance. In the past decade, great efforts have been spent striving for the longer underwater transmission distance while maintaining high-speed UWOC [9-19]. In long-distance and ultra-sensitive detection UWOC, on-off keying (OOK) modulation has outstanding anti-noise ability, while complicated modulation schemes, such as pulse position modulation (PPM) [10], discrete multi-tone (DMT) [13], pulse amplitude modulation (PAM) [17], and quadrature amplitude modulation (QAM) [19], may have potential of high data rate but involve complex digital signal processing (DSP) techniques. Usually, a collimated light beam with a small divergence from semiconductor laser diodes (LD) is utilized as the UWOC transmitter to propagate as long underwater distance as possible. Moreover, in order to achieve a longer transmission distance for high-attenuation underwater propagation, it is straightforward to increase the output power of UWOC transmitter. The high-power laser can be provided by beam combination of multiple LDs. Twelve bars of 450-nm laser are combined together and coupled into a fiber output of 1000 W [20]. In 2020, 3×1 fiber combiner is used to provide 2.4 W 450-nm laser for 100-

m/8.39 Mbps UWOC [16]. Also, blue-green beam combination by n×1 fiber combiner is demonstrated for wavelength division multiplexing (WDM) [21-24]. At the receiver side, because long-distance UWOC suffers from obvious aquatic attenuation and results in weak received optical power (ROP), highly sensitive photodetectors, such as single photon avalanche diode (SPAD) [25, 26], multi-pixel photon counter (MPPC) [10, 16], silicon photomultiplier (SiPM) [14], photomultiplier (PMT) [11, 17-19], etc., can provide satisfactory detection for weak light. Furthermore, coherent detection schemes have been demonstrated to enhance the sensitivity of receivers [27-30], relieving the heavy burden of UWOC detection due to the large underwater attenuation after long-distance transmission.

Table 1. Comparison of long-distance and high-speed UWOC in recent works.

| Year | Transmitter | Optical power | Photo-detector | Modulation | Underwater distance | Data rate | Reference |
|---|---|---|---|---|---|---|---|
| 2017 | 520-nm LD | 19.4 mW | APD | OOK | 34.5 m | 2.7 Gbps | [9] |
| 2018 | 450-nm LD | 0.17 mW | MPPC | 4-PPM | 46 m | 5 Mbps | [10] |
| 2019 | RGB LD | 5 W | PMT | OOK | 120 m | 20 Mbps | [11] |
| 2019 | 520-nm LD | 7.25 mW | APD | OOK | 100 m | 500 Mbps | [12] |
| 2019 | 452-nm LD | 12.8 mW | APD | DMT | 55 m | 5.6 Gbps | [13] |
| 2020 | 520-nm LD | 10 mW | SiPM | OOK | 40 m | 1 Gbps | [14] |
| 2020 | 532-nm LD | 1.4 W | APD | OOK | 100 m | 100 Mbps | [15] |
| 2020 | 450-nm LD | 2.4 W | MPPC | OOK | 100 m | 8.39 Mbps | [16] |
| 2021 | 450-nm LD | 0.3 W | PMT | PAM4 | 200 m | 500 Mbps | [17] |
| 2022 | 450-nm LD | 15 mW | PMT | OOK | 100 m | 3 Gbps | [18] |
| 2023 | 450-nm LD | 0.2 W | PMT | 32-QAM | 90 m | 660 Mbps | [19] |
| 2024 | 452-nm LD | 5 W | SiPM | OOK | 250 m | 1 Gbps | **This work** |

However, nonlinear impairments caused by bandwidth-limited high-power transmitter and sensitive receiver severely degrade the data rate of long-distance UWOC. To increase the data rate, equalization technique is employed to mitigate the inter-symbol-interference (ISI) generated from the limited bandwidth of the link [31, 32]. However, in principle, the linear equalizer cannot solve the nonlinearities of the link which mainly generated from the high-power light source and ultra-sensitive photodetector previously discussed for UWOC since that of electro-opto power conversion is a nonlinear function, and the limited bandwidth will introduce memory effect that further degrades the system performance [33]. In order to solve nonlinearity, researchers try to use Volterra nonlinear equalizer to mitigate nonlinear distortions [34, 35]. Specifically for UWOC system, in [17, 18], Volterra equalizer is used to achieve 500 Mbps/200 m and 3 Gbps/100 m UWOC performance respectively. Volterra equalizer used high-order kernels to solve the system nonlinearities. Other nonlinearity compensate technique includes digital back-propagation (DBP) [36], perturbation-based compensation [37], and nonlinear Kalman filter [38]. However, the formentioned equalization technique is expert-knowledge based, many nonlinearities such as modulation nonlinearity together with square law detection, non-linear power amplifiers (PAs), finite resolution quantization that can only be approximately captured by such models while is difficult to compensate with expert-knowledge based equalization technique. In order to solve this problem, DSP algorithms based on artificial neural network have been proposed [39, 40]. For LED-based link, in [41], Artificial Neural Network (ANN) equalizer significantly outperforms conventional nonlinear equalizer. Implemented in different optical communication systems, these ANN-based equalizers have not only reached lower bit error rate (BER), but also shown excellent capability of mitigating nonlinearity. However, one of the prominent challenges for the ANN-based equalizer is the high complexity that limit its application in high-speed real-time communication link. The complexity compared to the adaptive linear equalizer is dramatically increased when using multiple neurons and multiple layers. In this paper, to dramatically lower the complexity of the ANN equalizer, we applied a linear equalizer first in order to prune the input size of the ANN equalizer. The input size of the ANN equalizer here using memory lengths below order of 10 that achieves the saturated performance, when compared to the ANN equalizer using order of 50 in [39] and order of 80 in [40] that solely based on ANN.

Considering the implementation cost and reliability, the OOK modulation is more practical for the UWOC system with high-power transmitter and ultra-sensitive SiPM receiver. In this paper, a high-power UWOC transmitter with the typical output power of 26.2 W is realized by beam combination of 8-channel cascaded blue LDs and a 250-meter long underwater transmission channel is reliably constructed. Meanwhile, a sensitive SiPM detector is used for weak light detection at the receiver side. Then, with directly OOK-modulated LDs, we successfully achieve 1-Gb/s data transmission over the

250-m UWOC system by using combined linear equalizer and low-complexity ANN equalizer. Volterra and Linear equalizer based link performances are also compared. Experimental results demonstrate that the linear equalizer can support up to 500 Mbps and Volterra equalizer is able to support up to 750 Mbps. To the best of our knowledge, this is the first demonstration of Gigabits-per-second-level data transmission over >250-meter long distance for UWOC system.

## 2. Principle

The limited bandwidth of the communication link introduces ISI, which can significantly degrade the system performance. In addition, nonlinearities arise from both the light source and the optical detector. To mitigate these issues and enhance link performance, both linear and nonlinear equalizers are required. This section introduces three different types of equalizers.

### 2.1 Linear equalizer

Figure 1 illustrates the structure of a linear equalizer. The coefficient vector w(.) represents the parameters of the equalizer, which are trained using the least mean square (LMS) algorithm. The output of the linear equalizer can be expressed as:

$$y(n) = \sum_{m=0}^{M} x(n-m)w(m) \qquad (1)$$

where x (n) is the received signal, w(m) is the tap coefficient of the linear equalizer and y(n) is the recovered output signal. Fig. 1 depicts the fundamental structure of the linear equalizer. This equalizer is employed to mitigate the ISI caused by the link. The output of linear equalizer is then down sampled and fed into the ANN equalizer.

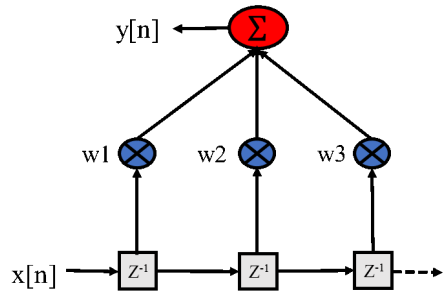

**Fig. 1.** Linear equalizer structure.

### 2.2 Volterra equalizer

The basic structure of a Volterra equalizer is shown in Fig. 2 and expressed in:

$$y(n) = \sum_{m=0}^{M-1} x(n-m)w1(m) \\ + \sum_{m1=0}^{M1-1}\sum_{m2=m1}^{M1-1} x(n-m2)x(n-m1)w2(m2,m1) + \sum_{m3=0}^{M1-1} x(n-m3)^3 w \qquad (2)$$

where *y[n]* is the recovered output signal, *x(n)* is the received signal, *w1* is the tap coefficient of linear equalizer which has the tap length of M, *w2* (.) is the tap coefficient of the second order kernel which has the length of $C_{M1}^2$+M1, $C_{M1}^2$ represents the combinational number, *w3* is the tap coefficient of the third order kernel which has the length of M3. It is evident that the Volterra equalizer encompasses a linear equalizer, while the second- and third-order kernels are introduced to address the nonlinearities presented in the system. The tap coefficients of the equalizer are trained using the LMS algorithm, which is also employed for training the linear equalizer. where *y[n]* is the recovered output signal, *x(n)* is the received signal, *w1* is the tap coefficient of linear equalizer which has the tap length of M, *w2* (.) is the tap coefficient of the second order kernel which has the length of $C_{M1}^2$ +M1, $C_{M1}^2$ represents the combinational number, *w3* is the tap coefficient of the third order kernel which has the length of M3. It is evident that the Volterra equalizer encompasses a linear equalizer, while the second- and third-order kernels are introduced to address the nonlinearities presented in the system. The tap coefficients of the equalizer are trained using the LMS algorithm, which is also employed for training the linear equalizer.

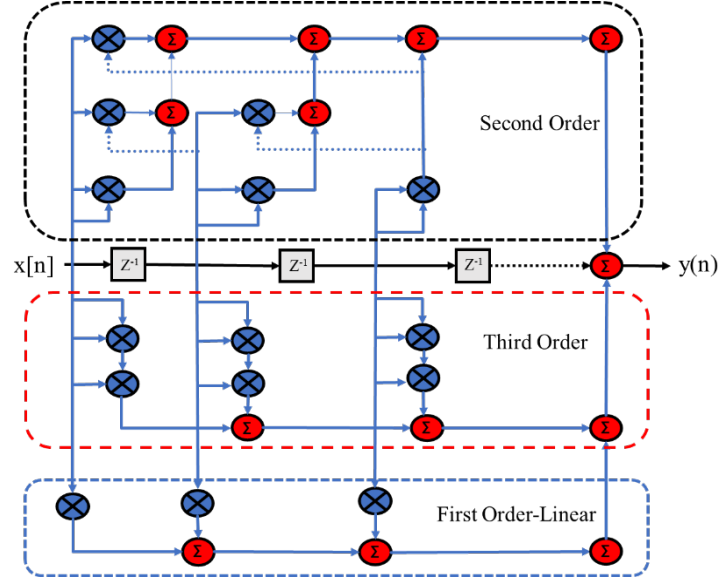

**Fig. 2.** Volterra equalizer structure.

*2.3 ANN equalizer*

The architecture of the artificial neural network employed in this study is illustrated in Fig. 3. The multiple-layer structure for deep-learning network has been discussed to mitigate nonlinear distortion for SPAD-based UWOC [25]. Here, a low-complexity artificial neural network with two hidden layers is utilized. The activation function in the hidden layers is the sigmoid function, while the output layer uses a linear activation function. The output of each layer, including the first hidden layer, the second hidden layer, and the output layer, is expressed as follows:

$$y1(z) = \sigma[\sum_{m=0}^{M} x(n-m)w1(z,m)] \quad (3)$$

$$y2 = \sigma[\sum_{q=0}^{Q} y(z-q)w2(q)] \quad (4)$$

$$y3 = w3y2 + b \quad (5)$$

where x(n) is the output of the linear equalizer after down sampling, *w1* is the coefficient matrix referring to first layer of neurons and *w2* is the coefficient matrix for the second layer. In this study, only a single neuron is employed in the second hidden layer, a choice that will be further elaborated upon in the experimental section.

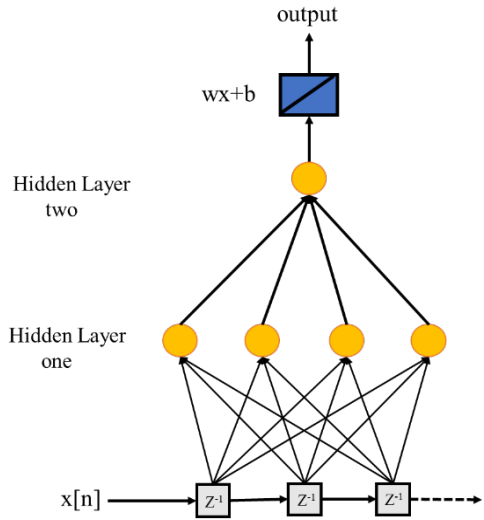

**Fig. 3.** ANN equalizer structure.

## 3. Experimental setup

### 3.1 Design for UWOC transmitter and receiver

A high-power 450-nm blue light laser serves as the transmitter for long-distance UWOC. The internal structure of the laser source is depicted in Fig. 4(a). Figure 4(b) is the photo image of 8-channel cascaded laser when it operates in a typical condition. By serially driving 8 individual LDs and spatially combining them into a single beam, the beam is coupled into a multimode fiber and a typical laser output of 26.2 W can be obtained. Figure 4(c) presents the output spectrum of the combined beam, showing a central wavelength of 452 nm with a 10-nm full width at half maximum (FWHM) intensity. Figure 4(d) illustrates the voltage-current (U-I) and optical power-current (P-I) measurement curves of the high-power laser.

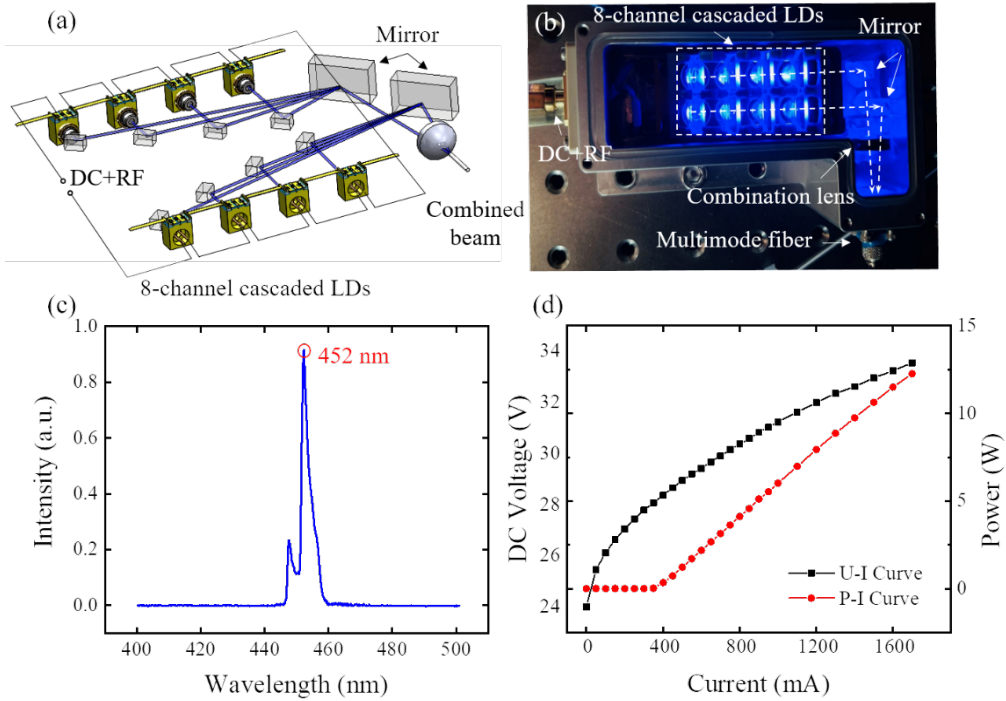

**Fig.4.** (a) Schematic diagram for beam combination of 450-nm, 8-channel cascaded high-power laser. (b) Photo image of 8-channel cascaded laser. (c) Measured output laser spectrum. (d) Measured U-I and P-I curves of the laser.

The UWOC receiver is designed to capture and focus the divergent light beam, which has propagated over a 250-m underwater channel, onto a SiPM detector. As shown in Fig. 5(a), the receiver is primarily composed of a variable-focus Cassegrain telescope system [42] with a 65 mm aperture, which collects and focuses the incoming light onto a camera sensor. The telescope's focal length is set to 250 m to ensure optimal focus on the camera's sensor. With a compact optical path and placement within a watertight-cabled assembly, the telescope design minimizes the space requirements. After passing through the variable zooming system, the focused beam is split into two paths via a coaxial optical setup, enabling signal detection by the SiPM and real-time imaging on the camera to confirm beam alignment. The optical distances from the reflective surface to both the detector and camera sensor are equal, maintaining consistent spot sizes across both sensors. The beam splitter has a 90:10 power ratio, directing most of the light energy to the SiPM to maximize the ROP for UWOC. Within the telescope, the incoming light is reflected and converged through the optical assembly, then directed through a 90:10 beam-splitting cube before focusing on the SiPM detector. The SiPM is onsemi 10035J series [43], with a 3 mm x 3 mm rectangular sensing area (9 mm²) and a high sensitivity to blue light of -40 dBm, enabling reliable detection of weak laser signals after 250 meters. To ensure accurate spot positioning, the splitter directs approximately 10% of the light to the camera, allowing the camera to capture the spot shape and assist in aligning it onto the detector. Ideally, camera-guided adjustment of the spot's size and position ensures maximum overlap with the detector's active area.

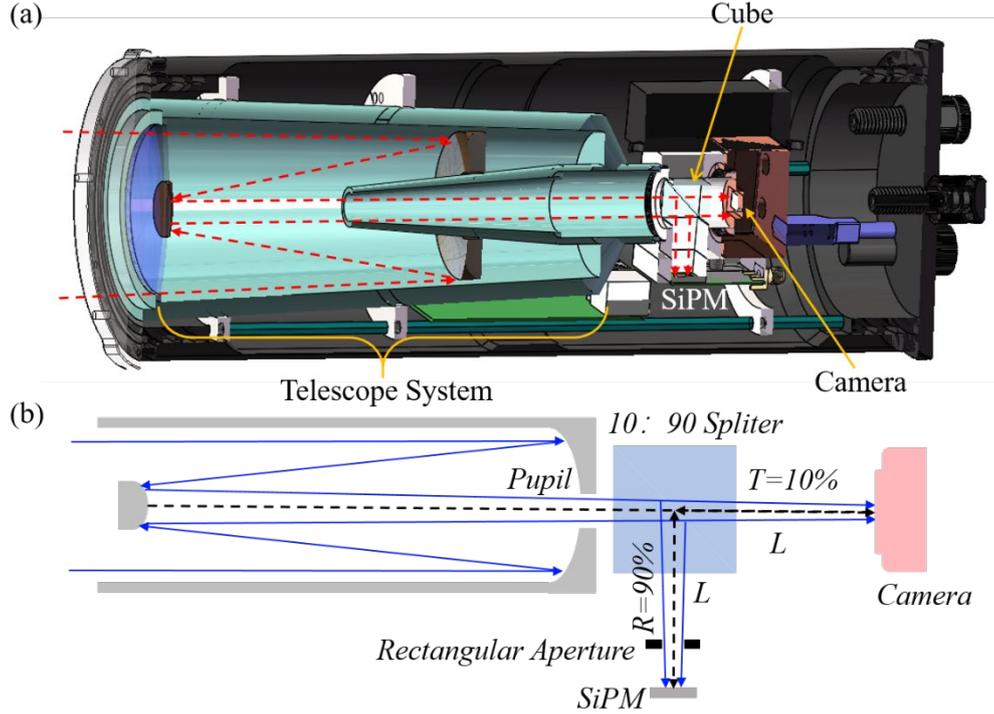

**Fig.5.** Structure of UWOC receiver. (a) physical assembly within the watertight enclosure. (b) schematic of the receiver, showing the telescope system and coaxial signal/imaging optical paths.

*3.2 UWOC system setup*

The experimental configuration for the 250-m blue light UWOC system is illustrated in Fig. 6(a), comprising 3 primary sections: transmitter, underwater channel, and receiver. For the transmitter, 300k bits of random binary data are encoded into OOK modulation and loaded into an arbitrary waveform generator (AWG) to generate the electrical signal. The electrical signal was loaded onto the RF port of bias-tee (Mini-Circuits, ZABT-80W-13-S+). Direct-circuit (DC) port of the bias tee is loaded to directly drive the 8-channel 450-nm laser diodes. The collimated laser beam emitted out of a collimator is injected into the underwater channel through a periscope. The spot of laser beam as going out of the periscope is shown in the inset on the left of Fig. 6(a), which is a uniform Gaussian distribution with a 25-mm diameter. Within the underwater channel, the beam is reflected for four times using four 50 cm×70 cm mirrors, extending the UWOC path to >250 m (50m×5). At the receiver end, the optical signal is detected by the SiPM enclosed in a watertight housing. The inset on the right of Fig. 6(a) shows the irregularly shaped beam spot after transmitting 250 m in the underwater channel as captured by the camera. Following underwater transmission, the light spot transforms from a circular Gaussian beam to an irregular distribution, becoming approximately 1 mm×2 mm in size. This irregularly beam shape typically results from turbulent-induced scintillation [44, 45]. Figure 6(b) shows the photo images of reflection pathway captured in 250-m underwater channels. The blue laser beam is emitted from the transmitter, marked as red arrows in Fig. 6(b), reflected for four times at 50 m, 100 m, 150 m, and 200 m, respectively, and finally reaches the receiver marked as white arrows in Fig. 6(b).

At the receiver end, the SiPM outputs an electrical signal and is recorded by the oscilloscope (Tektronix MS064B), with a 1 GHz analog bandwidth and sampling at 3.125 GSa/s. Subsequent offline signal processing, including synchronization, equalization, and BER computation, was performed using MATLAB. The oscilloscope captured 2 period of data pattern containing 600k bits and the first 300k bits are used to training the linear equalizer or Volterra equalizer. After training the linear equalizer, the other 300k bits are used to test the linear equalizer or Volterra equalizer. For the ANN testing, the output of the linear equalizer is down sampled and fed into the ANN equalizer. Then the 300k testing bits are divided into three classes: 50% for the training data set, 10% for the validation data set, and 40% which is 120k bits for the testing data set. Therefore, the testing data set is independent with the training data set in order to avoid the overfitting effect of the ANN equalizer. The system performance is evaluated and compared with three types of equalizers: linear equalizer, Volterra equalizer and the combined equalizer with linear + ANN.

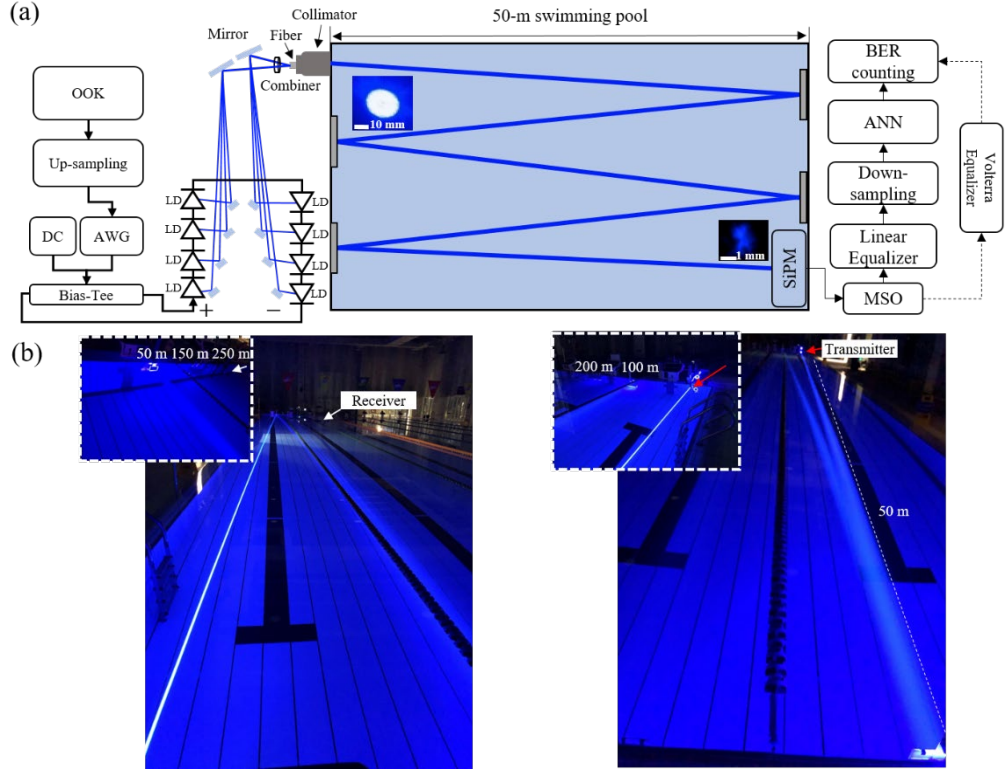

**Fig.6.** (a) Experimental setup of the proposed UWOC system. Insets: input (left) and output (right) light spots. (b) Reflection pathway of blue light within the 50-m swimming pool. Insets: spots reflected at 50 m, 150 m and receiver (left); spots at transmitter, and reflected at 100 m and 200 m (right). The UWOC transmitter is marked by red arrows and the UWOC receiver is marked by white arrows.

## 4. Experimental results and discussion

In this section, the performance of the low-complexity artificial neural network-based equalizer is systematically evaluated. Considering the signal-to-noise ratio (SNR) inherent in the long-distance UWOC link is low, the OOK modulation is employed for this demonstration. To provide a comparative analysis, the performance of linear and Volterra-based equalizers is also investigated. Firstly, measurements are conducted in an indoor experimental setup for optical wireless communication to optimize both the DC bias and RF peak-to-peak (Vpp) voltage for the link. The experimental configuration is shown in Fig. 7. The collimated light is directed through a spatial light attenuator and then split into two paths by 90:10 beam-splitting cube. 90% of the light is received by the SiPM, while the remaining 10% is captured by a camera or a power meter. By monitoring the size and position of the light spot on the camera, we can adjust the spot size irradiating the SiPM in real-time monitoring, ensuring complete coverage of the detector's target surface.

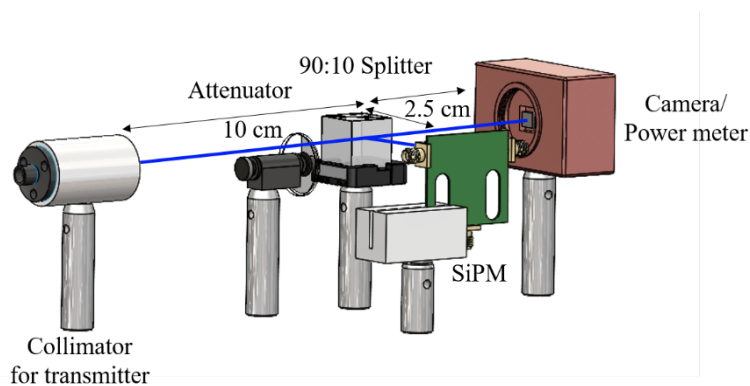

**Fig. 7.** Indoor experimental setup for optical wireless communications with high-power transmitter and SiPM receiver.

Fig. 8(a) further illustrates the BER performance as a function of varying DC bias voltages and RF peak-to-peak values, with a ROP of 150 nW. The BER is evaluated by recovering the signal using the linear equalizer. The results indicate that the optimal bias voltage is 27.95 V, with a peak-to-peak voltage of 6 V, which yields the most favorable performance. With the DC bias of 27.95 V, the output power of combined LDs approach ~5 W. Fig. 8(b) presents the BER variations of 1-Gb/s data transmission as ROP decreases from about -37 dBm to below -40 dBm. The results clearly indicate that the receiver sensitivity reaches -40 dBm, which corresponds to the hard-decision forward error correction (HD-FEC) limit.

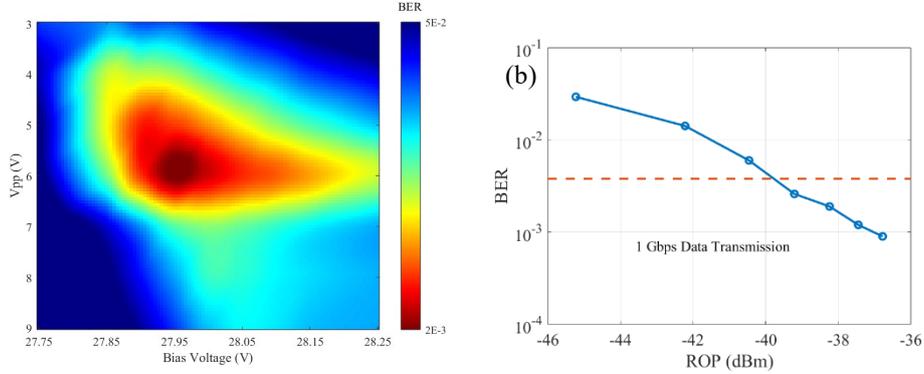

**Fig. 8.** (a) Different DC Voltage and RF peak-to-peak value versus BER performance (b) BER plot under different ROP at 1-Gb/s data transmission.

Fig. 9 presents the BER results for 250-m UWOC tests in a 50-m pool following the applications of the linear equalizer, Volterra equalizer, or linear + ANN equalizer, at the optimal DC bias point and RF Vpp voltage (DC = 27.95 V, Vpp = 6 V). The real-time monitoring of the beam spot on receiver (right inset in Fig. 6(a)) ensuring the alignment of UWOC link and measurement of the ROP of 150 nW. The ANN-based equalizer achieves much lower BER at various data rates than the linear equalizer and Volterra equalizer. It successfully delivers 1-Gb/s data transmission over a 250-meter underwater distance with a BER of $3.4 \times 10^{-3}$, below the HD-FEC limit, which is the world record for UWOC distance at Gbps-level data rate to the best knowledge of all authors. In contrast, the linear and Volterra equalizers achieve data rates of 500 Mbps and 750 Mbps, respectively.

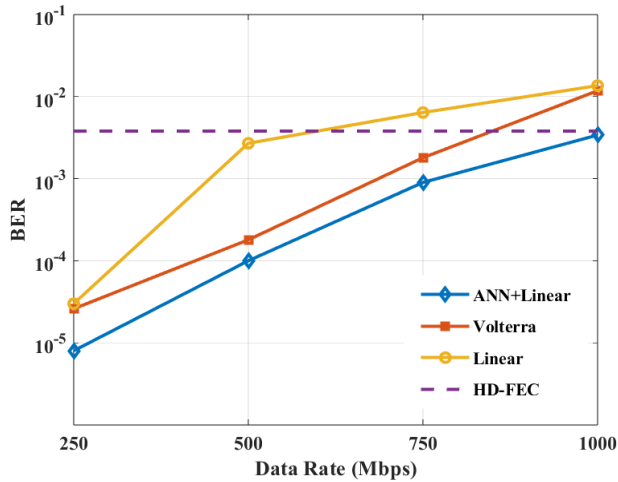

**Fig. 9.** BER for different type of equalizers under various data rates over 250-m UWOC results: the linear equalizer, Volterra equalizer and linear+ANN equalizer.

Fig. 10(a) depicts the memory length required for the linear equalizer under different data transmission rates. The saturated memory lengths are 300 taps, 200 taps, 100 taps, and 50 taps for data rates of 1 Gbps, 750 Mbps, 500 Mbps, and 250 Mbps, respectively. In the absence of equalization, the corresponding BER values are 0.3608, 0.2471, 0.1524, and 0.0675, respectively. These results clearly demonstrate that higher data transmission rates necessitate a longer memory length for the linear equalizer due to the increased severity of ISI. Building upon the linear equalizer's saturated memory lengths, the required memory length for the Volterra equalizer is also evaluated in Fig. 10(b). The Volterra memory length includes both second- and third-order kernel taps. The saturated performance of the Volterra equalizer is achieved with 4 taps, 8 taps, and 10 taps for 250 Mbps, 500 Mbps, and 750 Mbps data rates, respectively.

For 1-Gb/s data transmission, the performance improvement offered by the Volterra equalizer is marginal when compared to the linear equalizer. While the linear equalizer supports up to 500 Mbps, the Volterra equalizer can facilitate data transmission up to 750 Mbps, with the HD-FEC limit serving as the benchmark. However, both equalizers fail to support data transmission at 1 Gb/s.

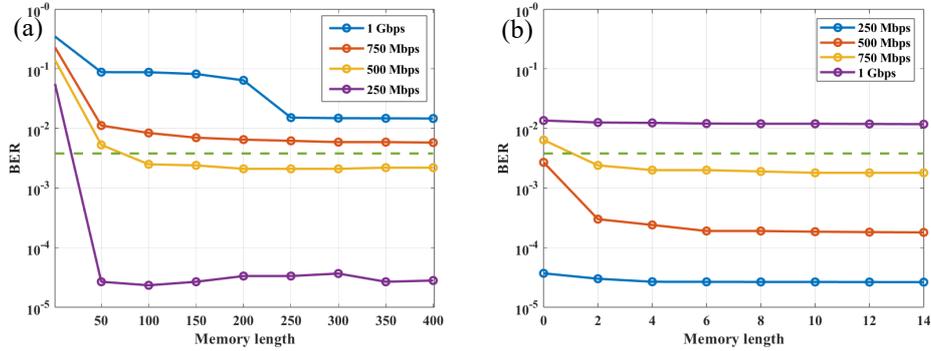

**Fig. 10.** 250-m UWOC data transmission: BER performance with different memory length for linear equalizer (a), and Volterra equalizer (b) under various data transmission rate.

Fig. 11(a) illustrates the BER as a function of the neural network input size and number of neurons in the first hidden layer, where the optimal input size is identified as 9 for ANN equalized 1-Gb/s data transmission. Increasing the input size beyond this point results in a slight deterioration in BER performance. This degradation is attributed to the limited size of the training dataset, which adversely impacts the convergence and generalization capacity of the artificial neural network during the training process. Fig. 11(b) also depicts the effect of varying the number of neurons in the second hidden layer on the system's BER performance. It is observed that, beyond a certain threshold, further increases in the number of neurons in the second layer lead to performance saturation, where no significant improvement in system performance is achieved.

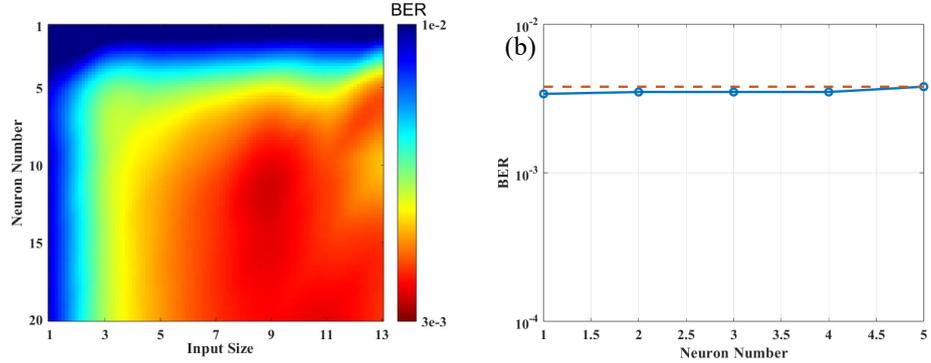

**Fig. 11.** 250-m UWOC system: BER performance with (a) different input size and neuron number in the first hidden layer, (b) different neuron number in the second hidden layer for ANN equalized 1-Gb/s data transmission.

Fig. 12 illustrates the training process of the ANN equalizer for 1-Gb/s underwater data transmission, showing the relationship between the input size and the mean squared error (MSE) for the testing dataset. In this experiment, the first hidden layer consists of 10 neurons, and the second hidden layer contains a single neuron. The ANN equalizer is trained using the Levenberg-Marquardt algorithm, known for its rapid convergence rate. As the input size increases from 1 to 9, a corresponding reduction in the MSE is observed. However, further increasing the input size beyond 9, specifically to 11, does not result in a further improvement in MSE.

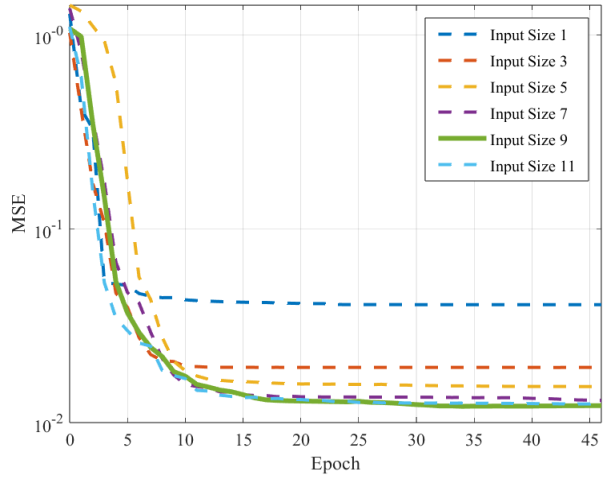

**Fig. 12.** 250-m UWOC system: ANN training process for different input size number for 1-Gb/s data transmission.

Fig. 13 illustrates the training process for data transmission rates of 250 Mbps, 500 Mbps, 750 Mbps, and 1 Gbps. The dataset is partitioned into three subsets: training, validation, and testing. The training set is utilized to optimize the coefficients of the ANN equalizer. The validation set is employed to monitor the equalizer's performance and determine when to halt the training process to prevent overfitting. The testing set is independent of both the training and validation sets, serving solely to assess the system's performance and calculate the BER. The MSE for the testing set stabilizes at $1\times10^{-5}$, $4.4\times10^{-4}$, $2.5\times10^{-3}$, and $1.2\times10^{-2}$ for data rates of 250 Mbps, 500 Mbps, 750 Mbps, and 1 Gbps, respectively. The MSE for the training set is observed to be the lowest, and a result of overfitting appears as the training process progresses. However, this overfitting effect does not translate into improved performance on the testing set.

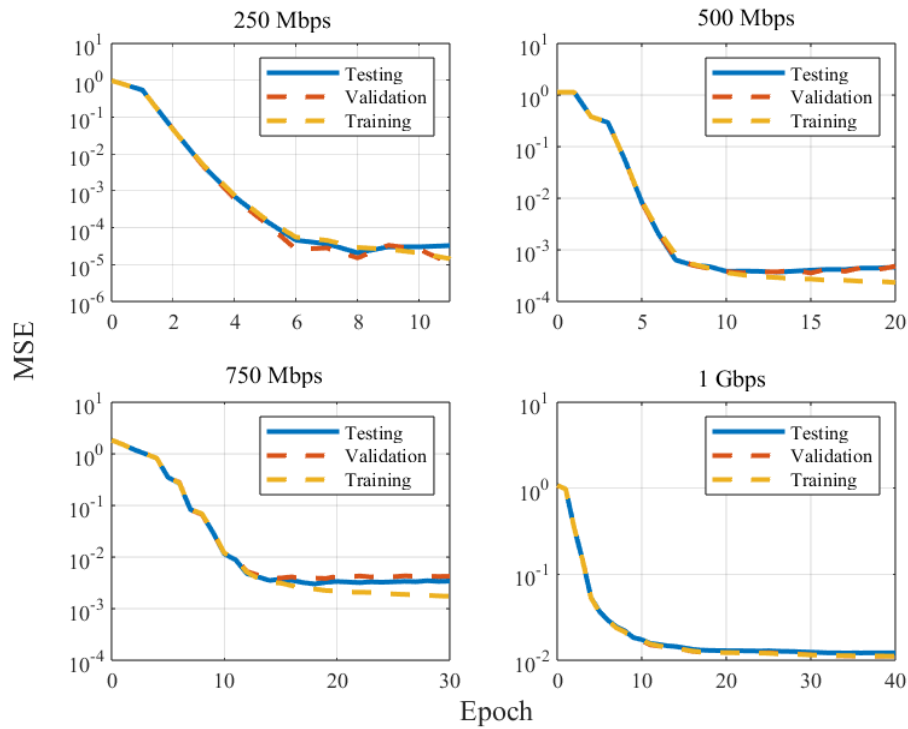

**Fig. 13.** 250-m UWOC system: ANN training process for various data transmission rate.

Fig 14 presents eye diagrams for 10k symbols across three types of equalization at various data rates (250 Mbps, 500 Mbps, 750 Mbps, and 1 Gbps). Among the three equalizers, the ANN exhibits the clearest eye diagram. The Volterra equalizer demonstrates superior eye quality compared to the linear equalizer at 250 Mbps, 500 Mbps, and 750 Mbps. However, both the linear and Volterra equalizers fail to achieve an open eye diagram at the 1 Gbps data rate. It is important to note that the ANN equalizer functions similarly to a decision circuit, owing to the characteristics of the sigmoid activation function. When the input to the ANN falls within a specific range, the output tends to stabilize at a fixed value, resulting in a significantly clearer eye diagram than the other two equalizers. In contrast, the eye diagrams for the linear and Volterra equalizers are degraded by both ISI and noise, leading to less distinct eye openings.

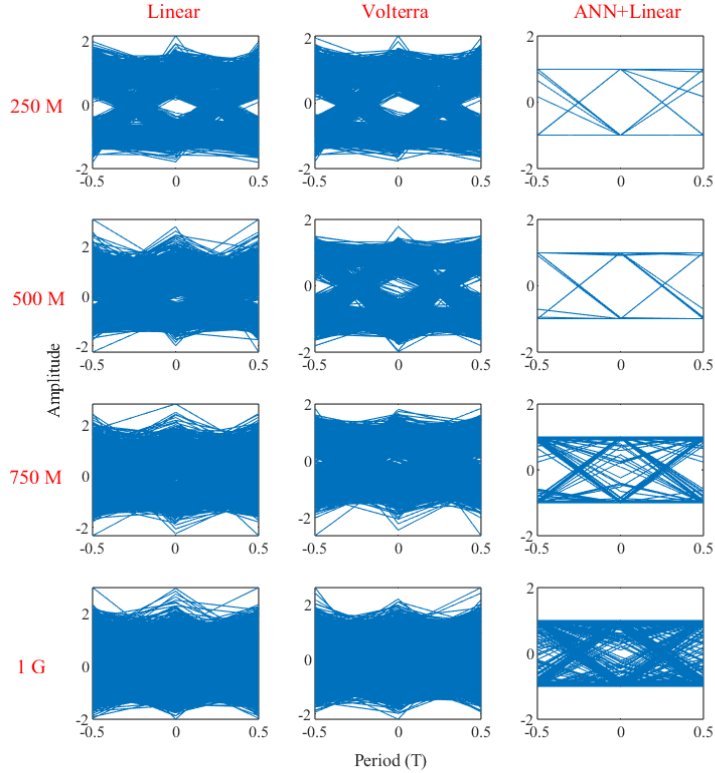

**Fig 14.** Eye diagrams under various data transmission rate for 250-m UWOC system.

## 5. Conclusion

In this paper, we propose a low-complexity ANN equalization scheme for long-distance UWOC systems. To lower the complexity of the ANN equalizer, a linear equalizer is applied first in order to prune the input size of the ANN equalizer. The optimal input size of the ANN equalizer is identified as 9. And the ANN architecture consists of two hidden layers, with 10 neurons in the first layer and a single neuron in the second layer. The performance of the proposed ANN-based system is compared with that of systems employing Volterra and linear equalizers. We successfully demonstrate 1-Gb/s data transmission over a 250-meter underwater distance with a BER of $3.4 \times 10^{-3}$, which is below the HD-FEC limit. In comparison, the linear and Volterra equalizer-based systems achieve data rates of 500 Mbps and 750 Mbps, respectively. The data rate of Gbps transmission is already sufficient to most UWOC applications, and this demonstration of >250-m underwater distance in a swimming pool will foresee plenty of bright future applications in long-distance and high-speed UWOC.

**Funding.** National Natural Science Foundation of China (62175120); Guangdong Basic and Applied Basic Research Foundation (2021B1515120084); Shenzhen Science and Technology Program (JCYJ20220818103011023).

**Disclosures.** The authors declare that there are no conflicts of interest related to this article.

**Data availability.** Data underlying the results presented in this paper are not publicly available at this time but may be obtained from the authors upon reasonable request.